\documentclass[epj]{svjour}
\usepackage{graphics}
\begin{document}
\title{Ultrarelativistic Electron-Positron Plasma}
\author{Markus H. Thoma\inst{1} 
}                     % Do not remove
\institute{Max-Planck-Institut f\"ur extraterrestrische Physik, Gie{\ss}enbachstr., 
85748 Garching, Germany}
\date{Received: date / Revised version: date}
\abstract{
Ultrarelativistic electron-positron plasmas can be produced in 
high-intensity laser fields and play a role in various astrophysical 
situations. Their properties can be calculated using QED at finite 
temperature. Here we will use perturbative QED at finite temperature 
for calculating various important properties, such as the equation of 
state, dispersion relations of collective plasma modes of photons and 
electrons, Debye screening, damping rates, mean free paths, collision 
times, transport coefficients, and particle production rates, of 
ultrarelativistic electron-positron plasmas. In particular, we will 
focus on electron-positron plasmas produced with ultra-strong lasers.
\PACS{
      {52.27.Ep}{Electron-positron plasmas}   \and
      {11.10.Wx}{Finite-temperature field theory} 
     } % end of PACS codes
} %end of abstract
\maketitle
\section{Introduction}
\label{intro}
Plasmas, i.e. (partly) ionized gases, are considered to be the fourth state of matter after the 
solid, liquid, and gaseous states. The plasma state dominates in the observable Universe: 99\% 
of the visible matter is in the plasma state, namely in the form of stars and hot interstellar 
and intergalactic gases. Plasmas emit light due to the excitation of atoms and ions and 
recombination. Plasmas can be produced by high temperatures, such as in the sun or fusion reactors,
by electric fields (discharges), as used for illumination in neon tubes or in lightening, or
by radiation, such as in the Crab nebula where the pulsar in the center emits synchrotron radiation.

Plasmas can be classified according to various aspects: Relativistic plasmas, e.g.
the electron-positron plasma in a supernova explosion, are plasmas in which
the thermal energy $k_BT$ of the plasma particles is of the order of their rest mass energy $mc^2$
or larger. Quantum plasmas, e.g. the degenerate electron component in a white dwarf, are plasmas
in which the thermal de Broglie wave length $\lambda_B=h/(m v_{th})$ is of the order of the 
interparticle distance $d$ or larger. Here $v_{th}$ is the thermal velocity of the particles.
Strongly coupled plasmas, e.g. the ion component in white dwarfs, are plasmas in which the 
interaction energy between the particles is larger than their thermal energy. In non-relativistic 
plasmas this corresponds to a Coulomb coupling parameter $\Gamma_C=Q^2/(dkT)>1$.     

Plasmas in Nature, like in comets, in aurorae, in the corona of the sun, in lightening,
in flames, or in the sun, and in technology, like in discharges or fusion reactors,
cover a wide range of pressures and temperatures. All of these plasmas are non-relativistic,
classical, and weakly coupled systems. 

Electron-positron plasmas (EPPs) are created in the presence of strong electric or magnetic fields 
or extremely high temperatures, where massive pair production sets in. For example in a supernova
explosion temperatures up to $3 \times 10^{11}$ K corresponding to $k_BT\simeq 30\> {\rm MeV} \gg
2 m_ec^2 = 1.022\> {\rm MeV}$ will lead to an ultrarelativistic EPP. Also
in the vicinity of magnetars, i.e. neutron stars with magnetic fields $B>10^{14}$ G, and in
accretion disks around black holes EPPs show up.

Recently the possibility to create ultrarelativistic EPPs with high-intensity 
lasers ($I>10^{18}$ W/cm$^2$) have been discussed. For example two opposite laser pulses hitting a 
thin gold foil will heat up the electrons in the foil up to several MeV leading to pair creation
\cite{Shen}.

In the following we will discuss the properties of an ultrarelativistic EPP
using quantum field theory (QED) at finite temperature. We will follow closely the review article
\cite{Thoma1}. We will not discuss the production mechanism and equilibration of the EPP here.

\section{Field theoretic description of an electron-positron plasma}
\label{sec:1}

Throughout the paper we will use natural units, i.e. $\hbar =c=k_B=1$, as usual in quantum field 
theory, in which all units are given in powers of MeV. The conversion to conventional units can 
be achieved by $1=\hbar c=1.97 \times 10^{-13}$ MeV m from which $1\> {\rm MeV} = 1.60 \times  
10^{-13}\> {\rm J} {\hat =} 5.08 \times 10^{12}\> {\rm m}^{-1} {\hat =} 1.52 \times 10^{21}\> 
{\rm s}^{-1}$ follows. In these units the electron charge $e=0.3$ corresponding to a fine 
structure constant $\alpha =e^2/(4\pi)=1/137$.

\subsection{Equation of state}
\label{subsec:1}

We will start with the equation of state of an EPP and compute it under the following assumptions:

1. ultrarelativistic EPP, i.e. $T \gg m$, 

2. thermal and chemical equilibrium,

3. equal electron and positron density, i.e. vanishing chemical potential,

4. ideal gas, i.e. no interactions in the plasma,

5. infinitely extended, homogeneous, and isotropic EPP.

We will relax some of these assumptions in the following sections.
According to these assumptions the distribution function of the electrons and positrons
is given by the Fermi-Dirac distribution
\begin{equation}
n_F (p)=\frac{1}{e^{p/T}+1}
\label{e3}
\end{equation}     
and of the photons by the Bose-Einstein distribution
\begin{equation}
n_F (p)=\frac{1}{e^{p/T}-1},
\label{e4}
\end{equation}     
where the momentum $p$ is identical to the energy $E$ of the particles in the ultrarelativistic
case. It should be noted that the photons are in equilibrium with electrons and positrons under 
the above assumptions, i.e. the system is actually an electron-positron-photon gas.

The particle and energy density can be calculated by integrating over the distribution
functions. The particle number density of the electrons and positrons follows from integrating over
the Fermi-Dirac distribution as
\begin{equation}
\rho_e^{eq}=g_F \int \frac{d^3p}{(2\pi )^3} n_F(p)=\frac{3}{\pi ^2}\> \zeta (3)\> T^3=0.37\> T^3,
\label{e5}
\end{equation}     
where $g_F=4$ is the number of degrees of freedom corresponding to the electrons
and positrons in the two spin states. Assuming a temperature of $T=10$ MeV, we find
$\rho_e^{eq}=370$ MeV$^3 = 4.9 \times 10^{40}$ m$^{-3}$.

The photon density follows accordingly by integrating over the Bose-Einstein distribution with
$g_B=2$ degrees of freedom corresponding to the two polarization states as 
$\rho_{ph}^{eq}=(2/\pi ^2)\> \zeta (3)\> T ^3=0.24\> T^3$.

The energy density of the electron-positron-photon gas is obtained from
\begin{eqnarray}
\epsilon^{eq} & = & g_F \int \frac{d^3p}{(2\pi )^3} p\, n_F(p) + 
g_B \int \frac{d^3p}{(2\pi )^3} p\, n_B(p)\nonumber \\ 
& = & \frac{11\pi ^2}{60}\> T^4= 1.81\> T^4,
\label{e6}
\end{eqnarray}     
where the photons contribute 36 \% to the energy density. Here the Boltzmann law, 
$\epsilon^{eq}\sim T^4$, holds also for the fermions because we neglected their masses.

For $T=10$ MeV we
find $\epsilon^{eq}= 3.8 \times 10^{29}$ J m$^{-3}$. In a volume of $10^{12}$ m$^3$ 
(corresponding to the size of a neutron star) the total thermal energy of the EPP is
$3.8\times 10^{41}$ J, which corresponds 
to about 10\% of the entire energy (without neutrinos) released in a supernova type II explosion.
In a volume of 1 $\mu$m$^3$ there is still an energy of $3.8 \times 10^{11}$ J contained.

The Coulomb coupling parameter of the EPP, which is a measure for the non-ideal behavior
of a plasma \cite{Ichimaru}, is given by $\Gamma_C = e^2/(dT)$, where 
$d\simeq {\rho _e^{eq}}^{-1/3}=2.7 \times 10^{-14}$ m is the interparticle distance. 
For $T=10$ MeV we find $\Gamma_C = 5.3\times 10^{-3}$ which shows that the EPP is a weakly 
coupled plasma. Therefore the ideal gas results for the equation of state derived above are 
a good approximation. After all, interactions in the EPP play an important role, for example, 
for the collective behavior of the plasma as discussed in the next section and for equilibration 
of the plasma. Obviously, the interactions can be treated by perturbation theory.

\subsection{Collective Phenomena}
\label{subsec:2}

Collective effects in a plasma are associated with long-range interactions within the plasma.
Important examples are Debye screening and plasma waves. In non-relativistic ion-electron 
plasmas \cite{Lifshitz} plasmas these phenomena can be described by classical transport theory
(Vlasov equation). For example, the 
electron plasma frequency reads
\begin{equation}
\omega_{pl}=\sqrt{\frac{4\pi e^2\rho_e}{m_e}}
\label{e1}
\end{equation}     
and the Debye screening length due to the electrons in the plasma
\begin{equation}
\lambda_D=\sqrt{\frac{k_BT_e}{4\pi e^2\rho_e}},
\label{e2}
\end{equation}     
where $\rho_e$ is the electron number density, $T_e$ the temperature of the electron component,  
and $m_e$ the electron mass. In an ultrarelativistic plasma with $T \gg m$ the masses can be 
neglected and the important scales are the temperature $T$, called the hard scale, and the soft 
scale $eT$, which determines the collective phenomena as we will see below.

Interactions between relativistic electrons and positrons can be treated by using perturbative QED.
This corresponds to an expansion in the fine structure constant $\alpha$. Most conveniently Feynman
diagrams are considered from which via Feynman rules quantities such as scattering cross sections,
decay and production rates, or life times can be calculated directly. In an EPP the interactions 
take place in the presence of a heat bath. Hence we have to consider QED at finite temperature.
For this purpose the Feynman rules are extended to finite temperature which can be achieved 
by using the imaginary or real time formalism \cite{Kapusta,Landsmann}. The calculations are similar 
to the ones done already within the last 30 years for the properties of the quark-gluon plasma
using perturbative QCD at finite temperatures. As a matter of fact, many results from the quark-gluon
plasma (see e.g. Ref.\cite{Thoma2}) can be directly carried over to the EPP.

\begin{figure}
\resizebox{0.50\textwidth}{!}{\includegraphics{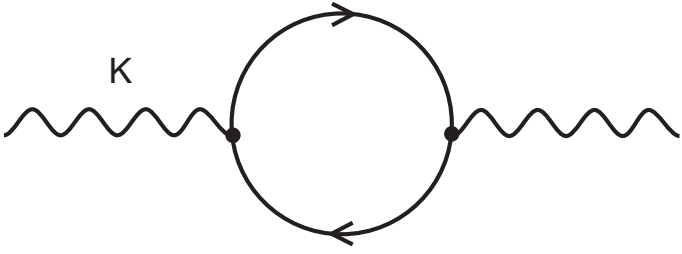}}
%\centerline{\psfig{figure=polarization.eps,width=10cm}}
%\vspace*{0.5cm}
\caption{One-loop polarization tensor.}
\label{fig1}
\end{figure}

An important quantity is the polarization tensor or photon self-energy.
The lowest order diagram for the polarization tensor is shown in Fig.1. Assuming the external
four momentum $K=(k_0 ,{\bf k})$ to be soft, i.e. the frequency $k_0=\omega$ and $k=|{\bf k}|$ 
to be much smaller than $T$, and the internal
loop momenta to be hard, an analytic 
result can be found using the real or imaginary time formalism \cite{Klimov,Weldon1}
\begin{eqnarray}
\Pi_L(\omega,k) &=& -3m_{ph}^2\> \left (1-\frac{\omega }{2k}\, \ln \frac{\omega + k}{\omega -k}
\right ),\nonumber \\
\Pi_T(\omega,k) &=& \frac{3}{2}\> m_{ph}^2\> \frac{\omega^2}{k^2}\> \left [1-\left (1-
\frac{k^2}{\omega^2}\right )\frac{\omega }{2k}\, \ln \frac{\omega + k}{\omega -k}
\right ]
\label{e10}
\end{eqnarray}
where $m_{ph} = eT/3$ is called the effective photon mass. For $T=10$ MeV we get 
$m_{ph} = 1$ MeV. 

The crucial 
quantity from which the collective phenomena are derived is the dielectric tensor 
relating the macroscopic electric field $D_i$ in the medium to the external field $E_i$ 
($i=x,y,z$), i.e. in momentum space
\begin{equation}
D_i(\omega ,{\bf k}) = \sum_j \epsilon_{ij}(\omega ,{\bf k})\> E_j(\omega ,{\bf k}).
\label{e7}
\end{equation}      
In the case of an isotropic medium it depends only on $\omega $ and on $k$ and has two independent components
\begin{equation}
\epsilon_{ij}(\omega,k) = \epsilon_T(\omega ,k)\> \left (\delta_{ij}-\frac{k_ik_j}{k^2}\right )
+ \epsilon_L(\omega ,k)\> \frac{k_ik_j}{k^2}.
\label{e8}
\end{equation}     

The dielectric tensor is closely related to the polarization tensor or photon self-energy by
\cite{Elze}
\begin{eqnarray}
\epsilon_L(\omega,k) &=& 1-\frac{\Pi_L (\omega,k)}{k^2},\nonumber \\
\epsilon_T(\omega,k) &=& 1-\frac{\Pi_T (\omega,k)}{\omega^2},
\label{e9}
\end{eqnarray}
where $\Pi_L$ and $\Pi_T$ are the longitudinal and transverse components of the polarization tensor,
respectively.

The dielectric functions following from 
(\ref{e9}) and (\ref{e10}) can also be derived from the classical Vlasov equation together with 
the Maxwell equations \cite{Silin}, since the high-temperature limit corresponds to the 
classical limit. 

The dispersion relations of collective plasma modes, i.e. propagation
of electromagnetic waves in the plasma, can be found by using the 
Maxwell equation, leading to 
\begin{eqnarray}
\epsilon_L(\omega,k) &=& 0,\nonumber \\
\epsilon_T(\omega,k) &=& \frac{k^2}{\omega^2}.
\label{e11}
\end{eqnarray}
Combining (\ref{e9}), (\ref{e10}), and (\ref{e11}) gives the dispersion relations 
$\omega_{L,T}(k)$ of the transverse
as well as longitudinal plasma waves as shown in Fig.2. The longitudinal branch, which 
does not exist in vacuum, is called plasmon as in the case of non-relativistic plasmas.
The transverse branch does not play a role in non-relativistic plasmas but is equally
important as the longitudinal one in relativistic plasmas. Both branches start at
the plasma frequency $\omega_{pl}=\omega_{L,T}(k=0)=m_{ph}$. 
Consequently the collective plasma waves have soft momenta of the order $eT$.
At high momenta $k\gg m_{ph}$ the transverse mode approaches the free dispersion 
$\omega_T = k$, corresponding to a real photon in vacuum, whereas the longitudinal
mode disappears, i.e. its spectral strength is exponentially suppressed.
For $T=10$ MeV we find $\omega_{pl}=1.5 \times 10^{21}$ s$^{-1}$. 

\begin{figure}
\resizebox{0.50\textwidth}{!}{\includegraphics{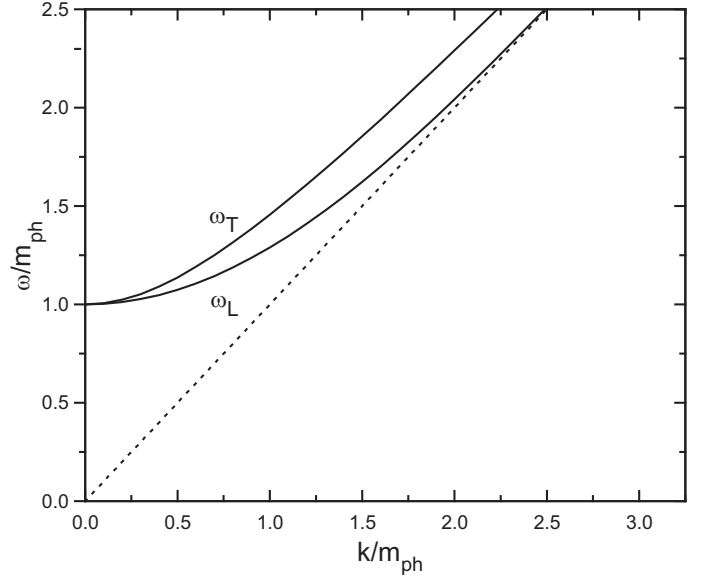}}
%\includegraphics[width=3in]{photon_dispersion.eps}
%\centerline{\psfig{figure=photon_dispersion.eps,width=11cm}}
%\vspace*{0.5cm}
\caption{Photon dispersion relation.}
\end{figure}

Another important quantity which can be derived from the polarization or dielectric tensor 
is the Debye screening length, entering the Yukawa potential of
a heavy, non-relativistic test charge in the EPP. The Debye screening length
is given by the static limit of the longitudinal component of the polarization tensor 
$1/\Pi_L(\omega =0)$, leading to 
$\lambda_D=1/(\sqrt{3}m_{ph})$, which is $1.1 \times 10^{-13}$ m for $T=10$ MeV. 

Finally from (\ref{e10}) we see that the polarization tensor and
the dielectric function become imaginary for $\omega^2<k^2$, i.e.
below the light cone $\omega =k$, corresponding to Landau damping. We also
observe that the plasma waves calculated at lowest order perturbation theory
are undamped since they are located at $\omega >k$. 

A complete new phenomenon that does not appear in non-relativistic plasmas is the
existence of fermionic plasma waves because all fermion masses are much too large
in the non-relativistic case. 
Their dispersion relations follow from the pole of the electron
propagator containing the electron self-energy. Using again the high temperature
approximation for the one-loop electron self-energy of Fig.3 leads to
($P=(p_0,{\bf p})$, $p=|{\bf p}|$)
\cite{Klimov,Weldon2}
\begin{equation}
\Sigma (P) = -a(p_0,p)\> P^\mu \gamma _\mu - b(p_0,p)\> \gamma _0
\label{e13}
\end{equation}
with
\begin{eqnarray}
a(p_0,p) & = & \frac {1}{4p^2}\> \left [tr(P^\mu \gamma _\mu\> \Sigma )
- p_0\> tr(\gamma _0 \> \Sigma )\right ]\; ,\nonumber \\
b(p_0,p) & = & \frac {1}{4p^2}\> \left [P^2\> tr(\gamma _0\> \Sigma )
- p_0\> tr (P^\mu \gamma _\mu \> \Sigma )\right ]\; ,
\label{2.28}
\end{eqnarray}
where the traces over the $\gamma $ matrices are given by
\begin{eqnarray}
tr(P^\mu \gamma _\mu \> \Sigma ) & = & 4\> m_F^2\; ,\nonumber \\
tr(\gamma _0\> \Sigma ) & = & 2\> m_F^2\> \frac {1}{p}\> \ln \frac
{p_0+p}{p_0-p}
\label{e14}
\end{eqnarray}
with the effective electron mass $m_F=eT/\sqrt{8}$, which is 1.1 MeV at $T=10$ MeV.

The full electron propagator in the helicity representation is given by \cite{Braaten1}
\begin{equation}
S^\star (P) = \frac {1}{2D_+(P)}\> (\gamma _0 - \hat p\cdot {\bf \gamma })
            + \frac {1}{2D_-(P)}\> (\gamma _0 + \hat p\cdot {\bf \gamma })
            \; ,
\label{e15}
\end{equation}
where
\begin{equation}
D_\pm (P) = -p_0 \pm p + \frac {1}{4p}\> \left [\pm tr (P^\mu \gamma _\mu
\> \Sigma ) - (\pm p_0 -p)\> tr (\gamma _0\> \Sigma )\right ]\; .
\label{e16}
\end{equation}

\begin{figure}
\resizebox{0.50\textwidth}{!}{\includegraphics{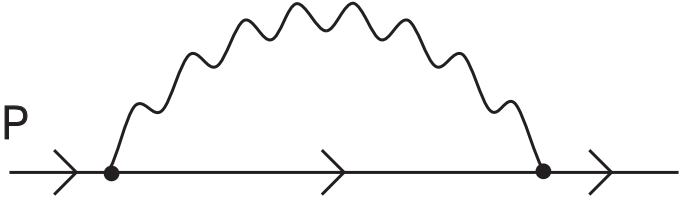}}
%\includegraphics[width=3in]{fermion.eps}
%\centerline{\psfig{figure=polarization.eps,width=10cm}}
%\vspace*{0.5cm}
\caption{One-loop electron self-energy.}
\label{fig3}
\end{figure}

The dispersion relations following from the pole of this propagator are shown in Fig.4. 
Again two branches show up, one with a positive ratio of the helicity to chirality 
($\omega_+$) following from $D_+=0$, the other one with a negative ratio ($\omega_-$) 
following from $D_-=0$, called plasmino \cite{Braaten1}. The plasmino branch 
$\omega_-$, which does not exist in vacuum, shows an interesting behavior, namely a minimum at
$k=0.41 m_F$, which may lead to van Hove singularities \cite{Braaten1,Peshier}.
Whether these van Hove singularities will lead to observable effects
in the EPP, e.g. in the electron spectrum, is a very interesting question which should 
be investigated in detail. It could open  the exciting possibility to observe a new 
collective plasma wave, the plasmino, experimentally in a laser induced EPP.
   
\begin{figure}
\resizebox{0.50\textwidth}{!}{\includegraphics{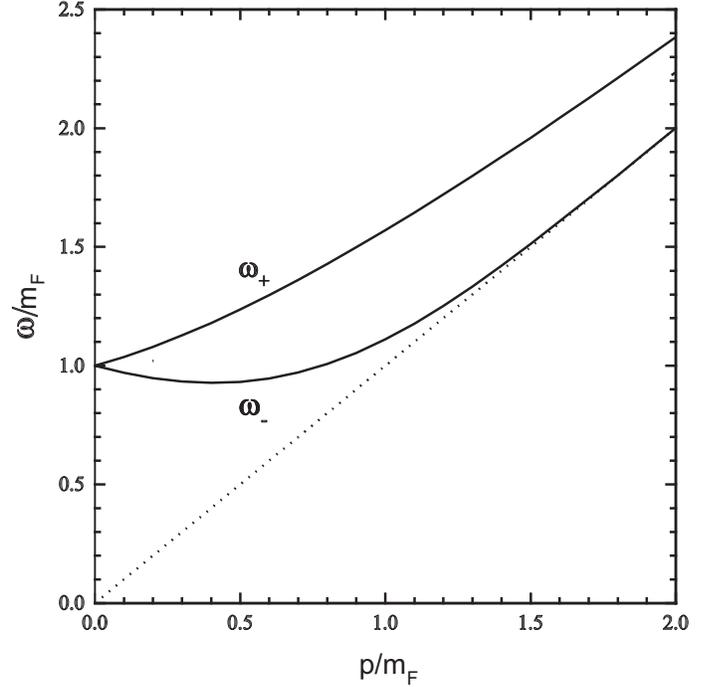}}
%\includegraphics[width=3in]{electron_dispersion.eps}
%\centerline{\psfig{figure=electron_dispersion.eps,width=10cm}}
%\vspace*{0.5cm}
\caption{Electron dispersion relation.}
\label{fig4}
\end{figure}

\subsection{Transport properties}
\label{subsec:3}

Now we want to consider the interaction and properties of particles in the plasma
with hard momenta, i.e. of the order of $T$ or larger. In particular we are interested
in damping and transport rates, mean free paths, collision times, energy losses 
of these particles and other transport properties such as the shear viscosity of the
EPP. 

It was shown by Braaten and Pisarski \cite{Braaten2} that a consistent treatment
of gauge theories such as QED at finite temperature, i.e.
for obtaining results that are gauge independent, infrared finite, and complete to 
leading order, require the use of an effective perturbation theory 
using resummed Green functions based on the hard thermal loop (HTL) approximation (HTL resummation technique).  

The damping rate of an electron or positron in the EPP is defined as the imaginary part of
the dispersion relation $\omega _{L,T}(p)$. To lowest order it follows from the elastic 
scattering diagram of
Fig.5. In the case of a hard electron or positron with momenta of the order of $T$ or higher 
it exhibits a quadratic infrared (IR)
divergence which can be reduced to a logarithmic one using a HTL resummed photon
propagator. This logarithmic singularity is expected to be cut-off by higher order contributions
leading to \cite{Thoma2}  
\begin{equation}
\gamma_e = \frac{e^2T}{4\pi}\ln\frac{1}{e}
\label{e17}
\end{equation}     
within logarithmic accuracy, i.e. the constant under the logarithm is not determined.
For $T=10$ MeV we obtain $\gamma_e = 86$ keV, which is much smaller than $\omega_{pl}=1$ MeV,
showing that the EPP is not overdamped. 

Physically more important are the transport rates $\Gamma $ which are related to the mean free
path and collision time of electrons and positrons in the EPP. They differ from the 
damping rate in cutting off the long range interactions with small scattering
angles $\theta$ by a factor $(1-\cos \theta)$ under the integral defining the
rate \cite{Lifshitz}. This leads to an improvement of the IR behavior (logarithmic
instead of quadratic singularity in perturbation theory) and a finite result
using the HTL method. Logarithmic divergent quantities can be treated consistently by splitting 
them into a soft part and a hard part, where the soft part is calculated using the HTL 
resummation technique \cite{Braaten3}.
For the transport rate we find to logarithmic accuracy 
\begin{equation}
\Gamma_e = \frac{e^4 T^3}{3 \pi s}\ln\frac{1}{e},
\label{e18}
\end{equation}     
where the Mandelstam variable $s=(P+K)^2$ is the square of the sum of the four momenta of the 
incoming particles in the scattering diagram of Fig.5. 
For thermal particles we replace $s$
by its thermal average $\langle s \rangle = 2\langle p\rangle_e \langle k\rangle_e \simeq 19.3 T^2$,
where $\langle p\rangle_e =\langle k\rangle_e = \epsilon_e^{eq}/\rho_e^{eq}= 3.11 T$.
Assuming again $T=10$ MeV, we get $\Gamma_e=0.54$ keV.

The mean free path $\lambda^{mfp}_e$ and collision time $\tau_e$ of the plasma particles 
(electrons and positrons)
are given by the inverse of the transport rate $1/\Gamma_e$,
leading to $\lambda^{mfp}_e=0.37$ nm and
$\tau_e = 1.2\times 10^{-18}$ s at $T=10$ MeV.  

\begin{figure}
\resizebox{0.30\textwidth}{!}{\includegraphics{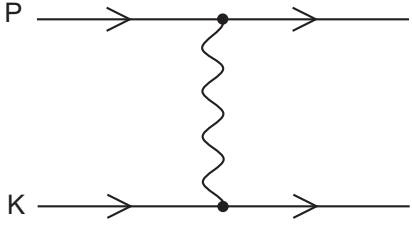}}
\caption{Lowest order diagram for electron-electron scattering.}
\label{fig5}
\end{figure}

In a non-relativistic plasma the shear viscosity can be estimated from elementary kinetic
theory as \cite{Reif} 
\begin{equation}
\eta _i = \frac {1}{3}\> \sum _i \rho _i\> \langle p_i\rangle \> \lambda^{mfp}_i
\label{e18a}
\end{equation}
where the sum is performed over the various components of the system.
In an relativistic plasma the coefficient 1/3 should be replaced by 4/15 \cite{deGroot}. 
Using the mean
free path following from (\ref{e18}), the density of (\ref{e5}), and the thermal 
momentum $\langle p\rangle_e =3.11 T$, 
the shear viscosity is given by (within logarithmic accuracy)
\begin{equation}
\eta_e = \frac{55.8 T^3}{e^4\ln (1/e)}.
\label{e19}
\end{equation}     
At $T=10$ MeV the shear viscosity coefficient is $\eta_e=7.9\times 10^{10}$ Pa s.

Another quantity of interest in a plasma is its stopping power or the energy loss of an energetic
particle in the plasma. There are two contributions, namely the energy loss by collisions
and the radiative one by bremsstrahlung. In a relativistic plasma the latter one becomes important.
The collisional 
energy loss is given by the mean energy transfer divided by the mean free path leading to 
\cite{Braaten4} 
\begin{equation}
\frac{dE}{dx} = \frac{1}{v} \int d\gamma \> \omega
\label{e19a} 
\end{equation}
where $v$ is the particle velocity, $\gamma$ the damping or interaction rate proportional 
to the plasma density
and the collision cross section, and $\omega$ the energy transfer from the
energetic particle to the plasma particle in the collision.  
Using for the collision cross section the lowest order diagrams in Fig.6, 
the collisional energy of a muon with mass $M$ in an EPP has been calculated 
by Braaten and Thoma \cite{Braaten4} applying the HTL resummation technique 
\begin{equation}
\frac{dE}{dx} = \frac{e^4 T^2}{24 \pi}\>  \left (\frac{1}{v} -
\frac{1-v^2}{2v^2}\> \ln \frac{1+v}{1-v} \right )\>  \left (\ln \frac{E}{M}
+ \ln \frac{1}{e} + A(v) \right ),
\label{e20}
\end{equation}
where $A(v)$ is a slowly varying function of the muon velocity $v$ between 1.3 and 1.5.

\begin{figure}
\resizebox{0.50\textwidth}{!}{\includegraphics{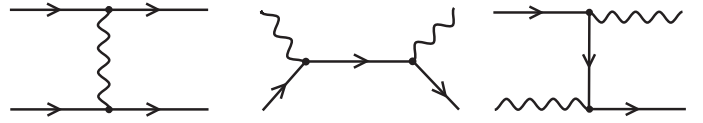}}
\caption{Diagrams defining the collisional energy loss.}
\label{fig6}
\end{figure}

The collisional energy loss of an electron with energy $E\gg T$ is approximately given by 
\cite{Braaten4}
\begin{equation}
\frac{dE}{dx}  = \frac{e^4 T^2}{48\pi}\> \ln \frac{15.3E}{e^2T}.
\label{e21}
\end{equation}
This leads to an energy loss of
200 MeV/nm for an electron or positron with an energy of $E=100$ MeV at $T=10$ MeV,
showing that such an electron is stopped (thermalized) within a fraction of a nanometer.
So far no calculations of the radiative energy loss in an EPP have been performed to our knowledge.

The damping rate of a photon in an EPP follows from the diagram in Fig.7, where a HTL resummed
electron propagator has to be used in case of soft momenta of the exchanged electron (positron).
In contrast to the electron damping rate, the photon rate is infrared finite using the HTL method 
due to the presence of an electron propagator in Fig.7 instead of the photon propagator in
Fig.5. Hence there is no need to cut off the long range interaction introducing a transport
cross section. The result for a photon with energy $E=p$ reads \cite{Thoma3}
\begin{equation}
\Gamma_{ph} = \frac{e^4 T^2}{64 \pi E}\ln\frac{3.88 E}{e^2T}.
\label{e22}
\end{equation}     
The mean free path and the collision time of photons in an EPP are given by $1/\Gamma_{ph}$.
For a thermal photon with the mean momentum 
$\langle p\rangle_{ph} =\epsilon_{ph}^{eq}/\rho_{ph}^{eq}=2.75 T$ at $T=10$ MeV 
the mean free path $\lambda^{mfp}_{ph}=0.28$ nm
and the collision time $\tau_{ph} =9.4\times 10^{-19}$ s follow. Actually the damping rate
given in (\ref{e22}) is a lower limit as higher order effects will enlarge it. As a matter of fact,
the photon production rate in a QGP, which is the inverse process of the damping rate \cite{Thoma3}, 
was shown to be about a factor of 2 larger taking bremsstrahlung into account \cite{Arnold1}.

\begin{figure}
\resizebox{0.50\textwidth}{!}{\includegraphics{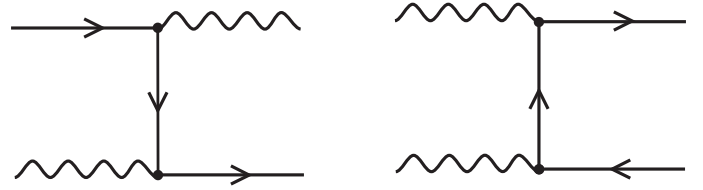}}
\caption{Diagrams defining the photon damping rate.}
\label{fig7}
\end{figure}

For the viscosity of the photon component using the above mean free path, the
photon density (see above), and the mean photon energy $\langle p\rangle_{ph} = 2.75 T$
we find  
\begin{equation}
\eta_{ph} = \frac{48.7 T^3}{e^4\ln (3.27/e)}
\label{e23}
\end{equation}     
corresponding to $3.5\times 10^{10}$ Pa s at $T=10$ MeV.
Hence the viscosity of the EPP $\eta =\eta_e+\eta_{ph}$ has similar contributions from the
electrons and photons.

A more advanced calculation of the total viscosity of the EPP based on the Kubo formula
yields within logarithmic accuracy \cite{Arnold2}
\begin{equation}
\eta = \frac{188 T^3}{e^4\ln (1/e)}.
\label{e23a}
\end{equation}
This result is about a factor of 1.5 larger than the one presented here based
on the elementary kinetic theory, which is typically valid within a factor of 2
\cite{Reif}.       

A summary of the QED results of the EPP properties discussed above is presented in the table below.

\bigskip

\noindent
%\small
\begin{tabular}{l|l|p{5cm}}
Quantity& Formula& Value at $T=10$ MeV\\ \hline
Electron-Positron Density& $\rho_e^{eq}=3/\pi ^2\> \zeta (3)\> T^3$& $4.9 \times 10^{40}$ m$^{-3}$\\
Photon Density& $\rho_{ph}^{eq}=2/\pi ^2\> \zeta (3)\> T^3$& $3.2 \times 10^{40}$ m$^{-3}$\\
Electron-Positron Energy Density& $\epsilon_e^{eq}=7\pi ^2/60\> T^4$& $2.4 \times 10^{29}$ J m$^{-3}$\\
Photon Energy Density& $\epsilon_{ph}^{eq}=\pi ^2/15\> T^4$& $1.4 \times 10^{29}$ J m$^{-3}$\\
Total Energy Density& $\epsilon^{eq}=11\pi ^2/60\> T^4$& $3.8 \times 10^{29}$ J m$^{-3}$\\
Thermal electron momentum& $\langle p\rangle_e = \epsilon_e^{eq}/\rho_e^{eq}=3.11\> T$& 31 MeV\\
Thermal photon momentum& $\langle p\rangle_{ph} = \epsilon_{ph}^{eq}/\rho_{ph}^{eq}=2.75\> T$& 28 MeV\\
Interparticle distance& $d\simeq {\rho_e^{eq}}^{-1/3}$& $2.7\times 10^{-14}$ m\\
Coulomb Coupling Parameter& $\Lambda = e^2/(dT)$& $5.3\times 10^{-3}$\\
Effective Photon Mass& $m_{ph}=eT/3$& 1 MeV\\
Plasma Frequency& $\omega_{pl}=m_{ph}$& $1.5 \times 10^{21}$ s$^{-1}$\\
Debye Screening Length& $\lambda_D=1/(\sqrt{3}m_{ph})$& $1.1 \times 10^{-13}$ m\\
Effective Electron Mass& $m_F=eT/(2\sqrt{2})$& 1.1 MeV\\
Electron Damping Rate& $\gamma_e = e^2T/(4\pi)\> \ln (1/e)$& 86 keV\\
Electron Transport Rate& $\Gamma_e = e^4 T^3/(3 \pi s)\> \ln (1/e)$& 0.54 keV for $s=19.3\> T^2$\\
Photon Damping Rate& $\Gamma_{ph} = e^4 T^2/(64 \pi E)\> \ln (3.88 E/e^2T)$& 0.70 keV for $E=2.75\> T$\\
Electron Mean Free Path& $\lambda_e^{mfp}=1/\Gamma_e$& 0.37 nm\\
Photon Mean Free Path& $\lambda^{mfp}_{ph}=1/\Gamma_{ph}$& $0.28$ nm\\
Electron Collision Time& $\tau_e=1/\Gamma_e$& $1.2\times 10^{-18}$ s\\ 
Photon Collision Time& $\tau_{ph} =1/\Gamma_{ph}$& $9.4\times 10^{-19}$ s\\
Electron Viscosity& $\eta_e = 55.8\> T^3/[e^4\ln(1/e)]$& $7.9\times 10^{10}$ Pa s\\
Photon Viscosity& $\eta_{ph} = 48.7\> T^3/[e^4\ln (3.27/e)]$& $3.5\times 10^{10}$ Pa s\\
Total Viscosity& $\eta =\eta_e+\eta_{ph}$& $(1.1 - 1.6)\times 10^{11}$ Pa s\\
Electron Energy Loss& $dE/dx=e^4T^2/(48\pi )\> \ln (15.3E/e^2T)$& 200 MeV/nm for $E=100$ MeV\\
\end{tabular}

%\normalsize 

\subsection{Non-equilibrium and finite chemical potential}
\label{subsec:4}

EPPs produced in  strong laser fields are probably not in complete equilibrium. For example,
it has been predicted by Shen and Meyer-ter-Vehn \cite{Shen} that a positron density of about 
$5\times 10^{28}$ m$^{-3}$ 
at a temperature of 10 MeV can be reached. This density deviates from the equilibrium
density (\ref{e5}) by 12 orders of magnitude. In the following we will therefore assume that
the EPP produced by lasers is in thermal but not in chemical equilibrium. Then we can replace the 
distribution functions for the electrons and positrons by Fermi-Dirac distributions multiplied by
a fugacity factor $\lambda $ describing the deviation from chemical equilibrium, 
$f_F(p)=\lambda n_F(p)$. This assumption has been used
for example for describing the chemical equilibration of the QGP in ultrarelativistic
heavy-ion collisions \cite{Biro}. The fugacity is given 
by the ratio of the experimental to equilibrium particle density, since the experimental
density follows from integrating over the non-equilibrium distribution, i.e.
\begin{equation}
\rho_{exp}=g_F \int \frac{d^3p}{(2\pi )^3} \lambda n_F(p)=\lambda \rho_{eq} \Rightarrow \lambda = 10^{-12}.
\label{e25}
\end{equation}     

Using the real time formalism, QED perturbation theory and the HTL method can also be extended to 
non-equilibrium situations like the one discussed above \cite{Carrington}. For example, the 
effective photon mass is given now by 
\begin{equation}
m_{ph}^2=\frac{4e^2}{3\pi^2} \int_0^\infty dp\> p\> f_F(p).
\label{e26}
\end{equation}     

\vspace*{12cm}

For $T=10$ MeV we then find for the non-equilibrium photon mass 
$m_{ph}^{noneq} = \sqrt{\lambda} m_{ph} = 1$ eV and the plasma frequency
$\omega_{pl}^{noneq}=1.5 \times 10^{15}$ Hz.
The Debye screening length in such an EPP is $\lambda_D=0.1$ $\mu$m. In order to speak of
a plasma with collective behavior its dimension $L$ should be much larger than
$\lambda_D$, i.e. at least of the order of 1 $\mu$m.

Furthermore, an anisotropic EPP can also be described by quantum field theoretic methods
\cite{Mrowczynski}. In this case instabilities can occur \cite{Randrup}.

Finally a possible difference between the positron density and the electron density can be 
treated by introducing a finite chemical potential $\mu$, i.e. using the distribution
\begin{equation}
n_F (p)=\frac{1}{e^{(p\pm \mu)/T}+1}
\label{e27}
\end{equation}     
for the electrons (negative sign) and positrons (positive sign).
Such a difference comes from the fact that the laser produced EPP is embedded in a hot electron
and cold ion background of the target. Therefore there will be an excess of electrons over positrons
in the hot, relativistic EPP. The methods described above,
such as the HTL resummation, can be generalized easily to this case \cite{Vija}. For example,
the energy density is given by
\begin{equation}
\epsilon^{eq} = \frac{11\pi ^2}{60}\> T^4 + \frac{1}{2}\> T^2\mu^2+\frac{1}{4\pi^2}\> \mu^4
\label{e28}
\end{equation}   
or the effective photon energy by 
\begin{equation}
m_{ph}^2=\frac{e^2T^2}{9} \left (1+\frac{3\mu^2}{\pi^2 T^2}\right ).
\label{e29}
\end{equation}

\subsection{Particle Production}
\label{subsec:5}

At high temperatures above 10 MeV also other particle species will be
produced, e.g. muons with a mass of $m_\mu = 106$ MeV. Their rate follows 
to lowest order from the diagram in Fig.8 (Born term).
We assume that $m_e \ll T\ll m_\mu$ holds. The first inequality 
implies that the electron mass can be put to zero and the latter inequality 
implies that muons are not equilibrated. Then the muon 
production rate to lowest order ($e^-e^+\rightarrow \gamma^* \rightarrow
\mu^-\mu^+$) is given by (for more details see Ref.\cite{Thoma1})
\begin{eqnarray}
&& \frac{dN}{d^4xd^4p}=\frac{\alpha ^2}{24\pi^4}\left (1+\frac{2m_\mu^2}{M^2}
\right ) \left (1-\frac{4m_\mu^2}{M^2}\right)^{1/2} \frac{T}{p} \nonumber \\ 
&& \frac{1}{\exp(E/T)-1}\ln \frac{1+\exp[-(E+p)/(2T)]}{1+\exp[-(E-p)/(2T)]},
\label{e24}
\end{eqnarray}
where $M^2=E^2-p^2$ is the invariant mass of the virtual photon 
$\gamma ^*$, $E$ its energy and $p=|\bf{p}|$ its momentum. 
Because of $M^2=E^2-p^2>4 m_\mu^2$ the rate is suppressed exponentially
for temperatures below $2m_\mu$.

In order to obtain the spectrum from this formula one has to integrate over 
the space-time volume, taking into account the space-time evolution 
by using, for example, a hydrodynamical model. The total muon yield then follows from 
integrating the spectrum over the energy and momentum of the virtual photon.

\begin{figure}
\resizebox{0.30\textwidth}{!}{\includegraphics{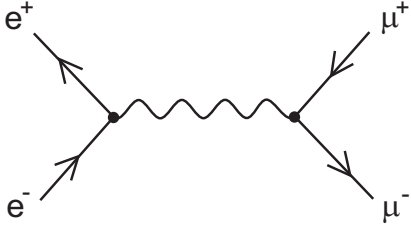}}
\caption{Lowest order contribution of the muon production.}
\label{fig8}
\end{figure}

At temperatures above 10 MeV also hadron production becomes important, in particular
pion production \cite{Kuznetsova}. 

\section{Conclusions}
\label{sec:3}

The aim of this presentation is the prediction of properties of ultrarelativistic
thermalized EPPs produced in laser fields or supernovae. The ultrarelativistic EPP is a weakly
coupled system. Therefore its equation of state can be described in first approximation
by an ideal ultrarelativistic gas. Interactions within the EPP can be described by 
perturbative QED at finite temperature. In this way collective phenomena (plasma waves,
Debye screening) and transport properties (damping rates, mean fee paths, relaxation times,
production rates, viscosity, energy loss) can be computed. A complete new phenomenon,
fermionic plasma waves (plasmino), which is absent in non-relativistic plasmas,
might be observable by van Hove singularities. The deviation from chemical equilibrium,
as expected for laser produced EPPs, can also be treated by extending QED to the
non-equilibrium case.   

\medskip

{\bf Acknowledgment:} I would like to thank D. Habs, J. Rafelski, I. Kouznetsova, and G. Moore
for helpful discussions and hints. The author acknowledges the support by the European Commission 
under contract ELI pp 212105 in the framework of the program FP7 Infrastructures-2007-1.

%\begin{figure}
%\resizebox{0.75\textwidth}{!}{%
%  \includegraphics{leer.eps}
%\vspace{5cm}       % Give the correct figure height in cm
%\end{figure}
%\begin{figure*}
%\vspace*{5cm}       % Give the correct figure height in cm
%\end{figure*}
%\begin{table}
%\begin{tabular}{lll}
%\hline\noalign{\smallskip}
%\end{tabular}
% Or use
%\vspace*{5cm}  % with the correct table height
%\end{table}
%
% BibTeX users please use
% \bibliographystyle{}
% \bibliography{}

\end{document}